\documentclass[aps, pre, twocolumn, notitlepage, showpacs, superscriptaddress, groupedaddress]{revtex4-1}

\usepackage{graphicx,epstopdf} 
\usepackage{amsmath}
\usepackage{graphicx}
\usepackage{caption}
\usepackage{subcaption}
\usepackage{amsfonts}
\usepackage{color}
\usepackage{dcolumn}  
\usepackage{bm}       
\usepackage{amssymb}  
\usepackage[breaklinks,colorlinks = true,linkcolor = red,urlcolor=blue,citecolor=red]{hyperref}
\usepackage{hyperref}
\usepackage{makecell}

\begin{document}

\newcommand{\angstrom}{\text{\normalfont\AA}}
\newcommand{\braket}[3]{\bra{#1}\;#2\;\ket{#3}}
\newcommand{\projop}[2]{ \ket{#1}\bra{#2}}
\newcommand{\ket}[1]{ |\;#1\;\rangle}
\newcommand{\bra}[1]{ \langle\;#1\;|}
\newcommand{\iprod}[2]{\bra{#1}\ket{#2}}
\newcommand{\logt}[1]{\log_2\left(#1\right)}
\def\cI{\mathcal{I}}
\newcommand{\cx}[1]{\tilde{#1}}
\newcommand{\nn}{\nonumber}
\newcommand{\la}{\langle}
\newcommand{\ra}{\rangle}
\newcommand{\p}{\partial}
\def\be{\begin{equation}}
\def\ee{\end{equation}}
\def\bea{\begin{eqnarray}}
\def\eea{\end{eqnarray}}

\newcommand{\eqa}[1]{\begin{align}#1\end{align}}
\newcommand{\mbf}[1]{\mathbf{#1}}
\newcommand{\iu}{{i\mkern1mu}}

\title{Kardar-Parisi-Zhang Scaling for an Integrable Lattice Landau-Lifshitz Spin Chain}

\author{Avijit Das$^1$, Manas Kulkarni$^1$, Herbert Spohn$^2$ and Abhishek Dhar$^1$}
\date{\today}							
	
\affiliation{ $^1$International Centre for Theoretical Sciences - Tata Institute of Fundamental Research, Bengaluru,  560089, India \\$^2$Zentrum Mathematik and Physik Department, Technische Universit\"{a}t M\"{u}nchen, Garching 85748, Germany}

\begin{abstract}
Recent studies report on anomalous spin transport for the integrable Heisenberg spin chain at its isotropic point.  Anomalous scaling is also observed in the time-evolution of non-equilibrium initial conditions, the decay of current-current correlations, and non-equilibrium steady state averages. These studies indicate a space-time scaling with $x \sim t^{2/3}$  behavior at the isotropic point, in sharp contrast to the ballistic form $x \sim t$ generically expected for  integrable systems.  In our contribution we study the scaling behavior for the integrable lattice Landau-Lifshitz spin chain. We report on equilibrium spatio-temporal correlations and   dynamics with  step initial conditions. Remarkably, for the case with  zero mean magnetization, we find strong evidence that  the scaling function is identical to the one obtained from the stationary stochastic Burgers equation, alias Kardar-Parisi-Zhang equation. In addition, we present results for the easy-plane and easy-axis regimes for which,  respectively,  ballistic and diffusive spin transport  is observed, whereas the energy remains ballistic over the entire parameter regime. 
\end{abstract}

\maketitle

\section{Introduction}
Classical Hamiltonian systems are usually classified as non-integrable and integrable, depending on whether they possess either a small or a macroscopically large number of conserved fields. More precisely, a Hamiltonian system with $N$ degrees of freedom is called integrable if one can find $N$ independent constants of motion -- otherwise, it is referred to as non-integrable. In general, one would expect  integrable and non-integrable systems to have drastically different  transport properties. Let us consider the example of a translation invariant one-dimensional  mechanical system with   $Q=\sum_{j=1}^N Q_j$ a conserved field  satisfying a local conservation law of the form $\partial Q_j/\partial t= J_{j}-J_{j+1}$, where $J_j$ is the corresponding local current. The corresponding dynamical equilibrium correlation function is defined by 
\begin{equation}\label{1.1}
 C(j,t)=\langle Q_j(t) Q_0(0) \rangle^{\mathrm{c}}_{\rm eq}, 
 \end{equation}
 where the average is over initial conditions chosen from the Gibbs equilibrium distribution and the superscript denotes the connected part of the correlator, defined as $\langle Q_j(t) Q_0(0) \rangle^{\mathrm{c}}_{\rm eq} := \langle ( Q_j(t) - \langle Q_0 \rangle_{\rm{eq}})( Q_0(0) - \langle Q_0 \rangle_{\rm{eq}}) \rangle_{\rm{eq}}$.  
 Since $Q$ is conserved, one expects a scaling form as
 \begin{equation}\label{1.2}
 C(j,t)= \chi (\Gamma t)^{-\alpha} f\big((\Gamma t)^{-\alpha}(j -ct)\big). 
 \end{equation}
 $\alpha >0$ is the scaling exponent, $c$ a potential systematic shift (the ``sound'' velocity), $\Gamma$ a model dependent parameter, and $f$ the scaling function normalized with total sum equal to $1$. Our 
 particular form ensures that $\sum_j  C(j,t) = \chi$ independent of $t$, and $\chi$ is the static susceptibility. 
 For integrable systems most commonly $\alpha = 1$ and  $c =0$, which is the ballistic behavior. The scaling function depends on  $Q$. 
 On the other hand, in non-integrable systems one often observes $\alpha = \tfrac{1}{2}$ with a Gaussian scaling function.  
 But also anomalous scaling with $\alpha = \tfrac{2}{3}$ has been discovered 
\cite{MendlSpohnPRL13, mendl14, SGDasPRE2014, Kulkarni2013, Kulkarni2015, spohn2014, SGDasArXiv2014, das2019}. Such differences between integrable and non-integrable systems are also observed in other transport simulations, for instance in the evolution of non-equilibrium initial conditions and in properties of boundary driven non-equilibrium steady states (NESS). Through generalized hydrodynamics much progress has been accomplished in the understanding of transport in integrable systems  \cite{spohn2018,doyon2018,PhysRevLett.121.230602,PhysRevX.6.041065,moore2017,PhysRevLett.117.207201}.

A surprising exception to the generic behavior has been discovered for spin transport in the integrable XXZ Heisenberg spin chain. The quantum $XXZ$ spin $\tfrac{1}{2}$ chain  is Bethe ansatz solvable for an arbitrary choice of the anisotropy parameter $\Delta$. The spectrum is gapless for $ |\Delta| \leq 1$ and gapped otherwise.  
A number of studies  find that, for zero $z$-magnetization, spin transport in this system is diffusive for $\Delta >1$, ballistic for $\Delta <1$ and anomalous at $\Delta =1$. First indications of this behavior came from the Drude weight for spin transport \cite{zotos1997,zotos1999,affleck2011}. Subsequent evidence was obtained in NESS studies at infinite temperatures \cite{prosen2009,znidaric2011}, in the form of equilibrium correlation functions \cite{Gemmer2009}, and in the evolution of quenched initial conditions \cite{moore2014}. There has been some understanding of the un-expected diffusive and anomalous regimes of spin transport using the GHD framework \cite{vasseur2018,PhysRevLett.121.230602}.

The goal of our contribution is to find out whether a similar pattern persists in corresponding classical models. One natural choice would be the classical Heisenberg model  on the one-dimensional continuum, also known as the Landau-Lifshitz model, which is integrable and has the same rotational symmetries as the XXZ chain \cite{c1,c2,c3,c4}. However numerical discretizations schemes might spoil integrability. Also the equilibrium states live on non-smooth spin configurations. For these reasons it is better to keep the underlying lattice, leading to the lattice Landau-Lifshitz (LLL) model, which in fact is non-integrable.  Recent work \cite{das2019} has explored spin and energy correlations. They turn out to be diffusive at high temperatures, while  anomalous features emerge at low temperatures. Fortunately one can adjust the coupling function between nearest neighbors on the lattice in such a way that the model is integrable and still has the usual rotational symmetries \cite{sklyanin1982,sklyanin1988,faddeev2007} . We will refer to this model as the integrable lattice Landau-Lifshitz (ILLL) system. The ILLL model has a parameter $\rho$, which plays the role of the anisotropy parameter $\Delta$ in the Heisenberg model, such that $\rho>0$ corresponds to easy-plane and $\rho<0$ corresponds to easy-axis, while $\rho\rightarrow 0$ is the isotropic case. At the isotropic point all three components of the total magnetization are conserved. For zero average magnetization the corresponding current correlations were studied in \cite{prosen2013}. Quite remarkably, the current correlation shows an  exponential decay for easy-axis ($\rho<0$) and hence a vanishing Drude weight and diffusive transport. Saturation to a non-zero value is observed for easy-plane ($\rho>0$), implying a finite Drude weight and ballistic transport. For the isotropic model an anomalous decay of the form $\sim t^{-\alpha}$ with $\alpha \approx  0.65$ is found. In our contribution we investigate the scaling properties of  equilibrium space-time correlations, for both spin and energy transport,  again for zero average magnetization. We confirm the scaling exponents from previous studies for spin transport in different parameter regimes. In addition, we determine the scaling functions: Gaussian for the case of diffusive transport in the easy-axis regime, while at the isotropic point, remarkably, we have a convincing fit to the Kardar-Parisi-Zhang (KPZ) scaling form $f_\mathrm{KPZ}$ [described later in Equation.~\eqref{eq.12}] . The energy transport remains ballistic in all regimes.

Returning to the quantum XXZ chain, in \cite{prosen2017,PhysRevLett.117.207201} the time evolution of the magnetization profile with a tiny initial  step is studied. As expected from linear response, it is observed that for $\Delta >1$ the magnetization profile has diffusive scaling $\alpha = \tfrac{1}{2}$, while, for $\Delta = 1$, the respective scaling  function is anomalous with $\alpha = \tfrac{2}{3}$.   In more recent work \cite{prosen2019}, it was found that the scaling function is related to the stationary KPZ equation. Such dynamical  properties live on the ballistic hydrodynamic scale and are implicitly  based on the assumption of local equilibrium. A similar reasoning can be applied to classical systems and generically one expects  to have comparable dynamical properties. Our study of the ILLL allows us to test more sharply whether such a conjecture holds. Indeed, we observe a clear KPZ scaling in the correlation functions. However, when investigating whether the  KPZ scaling  also holds for the evolution of the step-profile our data are too noisy for arriving at a definite conclusion.  In a recent complementary study \cite{gamayun2019,misguich2019}, the classical-quantum correspondence has  been analyzed in the context of the  continuum Landau-Lifshitz model, which is integrable, and the quantum XXZ model at zero temperature.

Our paper is organized as follows.  In Sec.~\ref{model}, we provide the details of the model, discuss the quantities of interest, and give a brief analysis using linear response theory. We also describe the numerical methods used.   Sec.~\ref{sec:eqsim} focuses on analyzing the equilibrium time correlations for all the three cases of the classical ILLL chain -- isotropic, easy-plane, and easy-axis regimes. In Sec.~\ref{sxjt}, we study the evolution of an initial step magnetization profile and its scaling.  We summarize our findings in Sec.~\ref{conc} along with an outlook. 

\section{The classical chain}
\label{model}
The classical ILLL \cite{sklyanin1982,sklyanin1988,faddeev2007} spin chain for $N$ spins, $\vec{S}_{j}, j = 1,...,N$, $|\vec{S}_{j}| =1$, is defined by the following Hamiltonian 
\begin{equation}\label{eq:1}
H = \sum_{j=1}^N   h(\vec{S}_j, \vec{S}_{j+1}),
 \end{equation}
 where the nearest neighbor interactions are given by 
\begin{eqnarray}\label{eq:2}
&&h(\vec{S},\vec{S'}) = -\log\big|\cos(\gamma S^{(z)})\cos(\gamma S'^{(z)}) \nonumber\\
&&\hspace{10pt}+ (\cot(\gamma))^2\sin(\gamma S^{(z)})\sin(\gamma S'^{(z)})\nonumber\\
&&\hspace{10pt}+ (\sin(\gamma))^{-2} G(S^{(z)}) G(S'^{(z)}) (S^{(x)}S'^{(x)} + S^{(y)}S'^{(y)})\big|, \nonumber\\
&&G(x) = \big(1-x^2\big)^{-\tfrac{1}{2}}\big(\cos(2\gamma x)-\cos(2\gamma)\big)^{\frac{1}{2}}.
\end{eqnarray}
$\gamma$  is the model parameter which can be either real or purely imaginary. Without loss of generality, we introduce the new parameter $\rho = \gamma^2$, $\rho \in \mathbb{R}$. The boundary conditions will be taken to be either periodic or open, depending  on the particular physical situation studied.  Easy-plane corresponds to $\rho>0$, easy-axis to $\rho <0$, while in the limit $\rho \to 0$ one obtains the isotropic interaction, 
 \begin{equation}\label{eq:3}
h\left(\vec{S},\vec{S'}\right) = -\log\left(1+ \vec{S}\cdot\vec{S'}\right).
 \end{equation}
Note that the $``-"$ sign in front of $h$ corresponds to the ferromagnetic interaction, whereas the positive sign will correspond to an anti-ferromagnetic interaction. In the present work we focus on the ferromagnetic Hamiltonian. For the anti-ferromagnetic case, the potential is not bounded from below and hence there would be equilibration problems at low temperatures. To see this, we note that for the general case with $h(\vec{S},\vec{S'}) = -J \log\big(1+ \vec{S}\cdot\vec{S'}\big)$, with $J >0$ ($J<0$) corresponding to ferromagnetic (antiferromagnetic) interactions, the  equilibrium state is given by,
\begin{equation}\label{eq:ES}
\prod_j \big(1 + \vec{S}_j\cdot\vec{S}_{j+1}\big)^{\beta J}.
\end{equation} 
For $J < 0$ this Boltzmann weight becomes unbounded as two neighboring spins point oppositely and can no longer be normalized once $\beta J \leq -1$.
Close to that value typically the chain will have long anti-ferromagnetic domains, which slow down the evolution.   A trace of this feature is still present at $\beta = 0$. 
Thus we find  for $\beta=0$ and  $J=1$ that after $10^6$ averages the data are still too noisy to pin down the tail behavior. More precise numerical data are achieved for $\beta = 1$, and we  use this value for all the simulations presented in this paper.   

The dynamics of spins is governed by Hamilton's equations of motion,
\begin{equation}\label{eq:4}
\frac{d}{dt}\vec{S}_j = \{\vec{S}_j,H\}= \vec{S}_j \times \vec{B}_j,\quad \vec{B}_j = -\nabla_{\vec{S}_j} H.
\end{equation}
We study transport in this model, through both equilibrium and nonequilibrium properties. \medskip \\
(i) In the equilibrium simulations we use periodic boundary conditions and compute spin and energy 
 space-time correlators defined by
\begin{eqnarray}\label{3.7}
C_\mathrm{ss}(j,t) &&=\langle S_j^{(z)}(t) S_{0}^{(z)}(0) \rangle^\mathrm{c}_\mathrm{eq},\nonumber\\
C_\mathrm{ee}(j,t) &&= \langle e_j(t) e_0(t) \rangle^\mathrm{c}_\mathrm{eq}
\end{eqnarray} 
where $e_j = h(\vec{S}_j,\vec{S}_{j+1})$  and the truncated average $\langle \dots \rangle^\mathrm{c}_\mathrm{eq}$ is taken with respect to the equilibrium distribution $e^{-\beta H}/Z$.\medskip\\
(ii) In the nonequilibrium simulations, we consider an open  XXX chain  initially prepared with a uniform temperature and a step in the magnetization.
More precisely, the initial state is $Z^{-1}\exp\big[-\beta \big(H-\sum_j h_j^{(z)}S_j^{(z)}\big)\big]$, where $h_j^{(z)} = -h_0$ for $j \leq 0$ and $h_j^{(z)} = h_0$ for $j > 0$. 
Of interest is the average magnetization  $s(j,t)=\langle S^{(z)}_j(t) \rangle_{h_0}$ at time $t $, where the dynamics is according to $H$ and the index recalls the dependence of the initial state on $h_0$. If the step is small, this average can be expanded in  $h_0$. The zeroth order vanishes and, using that $C_{\mathrm{ss}}(j,t) = C_{\mathrm{ss}}(-j,t)$, to first order one arrives at  
\begin{eqnarray}
s(j,t) &&= \beta \sum_i h^{(z)}_i C_{\mathrm{ss}}(j-i,t)\nonumber\\
&&=  \beta h_0\bigg(C(0,t) + 2  \sum_{i=1}^j C_{\mathrm{ss}}(i,t)\bigg)
\end{eqnarray}
for $j \geq 1$ with $s(- j +1,t) = - s(j,t)$.
As a  consequence, if $C_{\mathrm{ss}}(j,t)$ scales as in \eqref{1.2} with $c=0$, then $s(j,t)$ inherits the corresponding scaling. In the continuum form one arrives at
\begin{align}\label{eq:2.11}
s(x,t) = 2\beta h_0 \chi \int_0^{x/(\Gamma t)^\alpha} dx' f(x')
\end{align}
for $x\geq 0$. The derivative of the scaling function of $s$ yields the scaling function for $C$. For example, if $C_\mathrm{ss}$  has Gaussian scaling, then $s(x,t)$ would scale with the error function. Note that the scaling exponent remains unchanged.\medskip\\

{\it KPZ equation and scaling functions}. KPZ equation describes the surface growth under a random ballistic deposition. The height function $h(x,t)$ is governed by the 
Langevin equation,
\begin{equation}
\partial_t h = \tfrac{1}{2} \lambda (\partial_x h)^2 + \nu \partial_x^2 h + \sqrt{D} \eta,
\end{equation}
where $\eta$ is normalized space-time white noise.  The slope $u(x,t) = \partial_x h(x,t)$ is governed by the stochastic Burgers equation
\begin{equation}
\partial_t u +\partial_x\big( - \tfrac{1}{2} \lambda u^2 - \nu \partial_x u -\sqrt{D} \eta \big) =0.
\end{equation}
In the stationary state the mean of $u$ can be chosen to vanish and $x\mapsto u(x,0)$ is spatial white noise of strength $\chi = D/2\nu$. As shown in \cite{prahofer2004exact} the  two-point function of the stationary stochastic Burgers equation is given by
\begin{equation}\label{eq.12}
\langle u(0,0) u(x,t) \rangle \sim \chi(\Gamma t)^{-2/3} f_\mathrm{KPZ} \left((\Gamma t)^{-2/3}x\right).
\end{equation}
$\Gamma$ determines the non-universal time scale, which in the case of the Burgers  equation turns out to be $\Gamma= \sqrt{2} \lambda$. The scaling function $f_\mathrm{KPZ}(x)$ is positive, symmetric relative to the origin, and normalized to $1$. It looks like a Gaussian in bulk 
but has tails which decay as $\exp(-0.295 |x|^3)$, hence faster than a Gaussian. \medskip\\
{\it Simulation details}. We integrate the evolution equation \eqref{eq:4} using the adaptive Runge-Kutta method \cite{flannery1992numerical}. In some some regions in configuration space, the logarithmic interaction potential is very steep, and because of this the fixed step-size Runge-Kutta method turned out to be insufficient, especially at large times. One challenge is to keep the energy and the lengths of individual spins conserved during the numerical integration. Both these quantities dissipate quite a bit with time due to the accumulation of numerical errors. We give the input tolerance in the adaptive algorithm such that at the final time total energy remains conserved up to four decimal places and individual lengths of spins up to five decimal places. Total magnetization remains conserved well, up to 13 decimal places.
\begin{figure*}[]
\includegraphics[width=\textwidth]{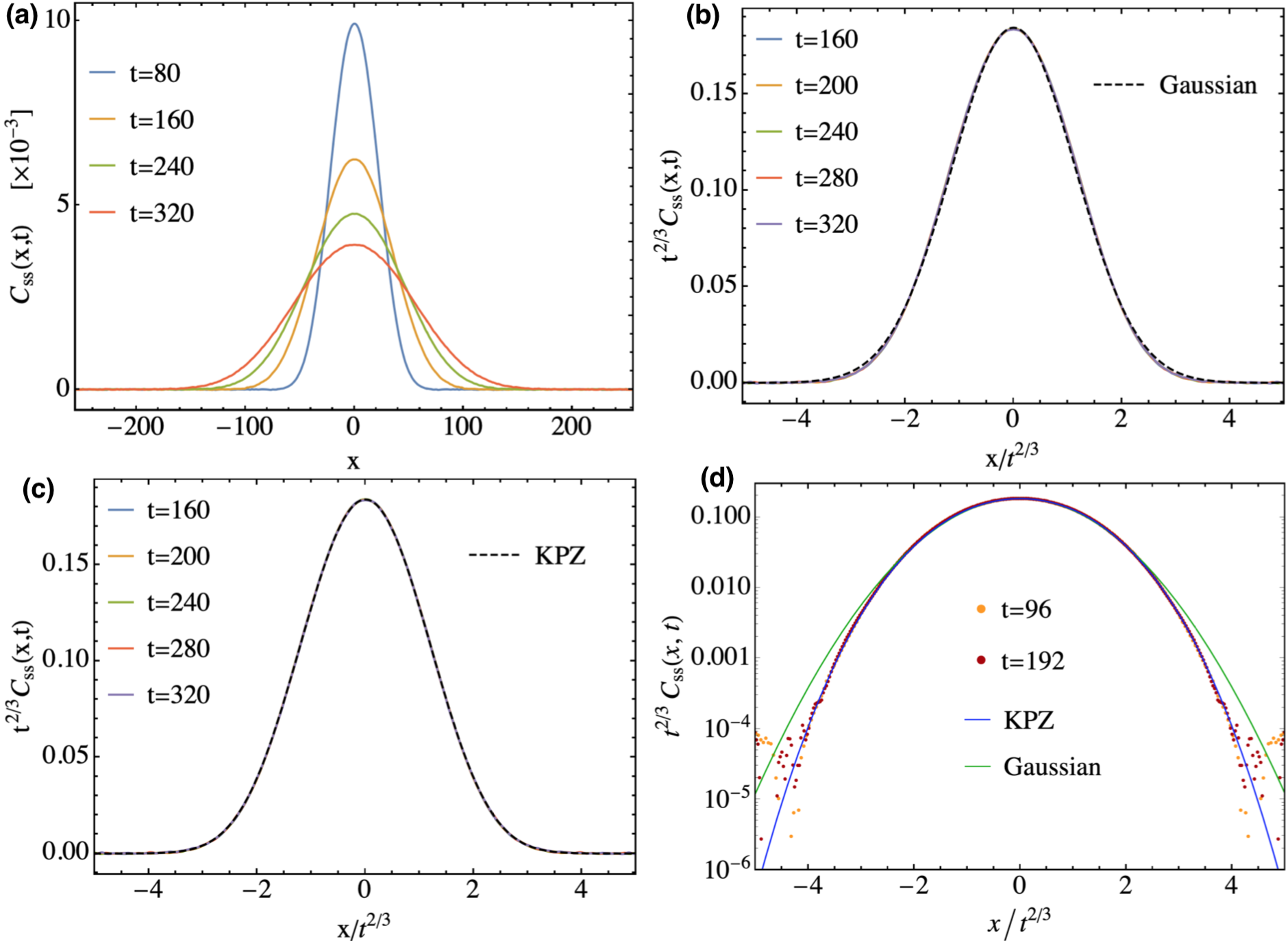}
\caption{(Isotropic regime) (a) Plot of the spin-spin correlation $C_\mathrm{ss}(x,t)$. And the same after a $t^{2/3}$ scaling with a fit to (b) Gaussian and (c) the KPZ scaling functions. In (d) we show the two fits compared to the data in logarithmic $y-$scale. This plot reveals that the KPZ scaling function offers a much better fit to the data. Parameter values: system size = $2048$, averaging over $\sim 10^6$ initial conditions and inverse temperature $\beta=1$.}
\label{fig:1}
\end{figure*}
\begin{figure*}[]
\includegraphics[width=\textwidth]{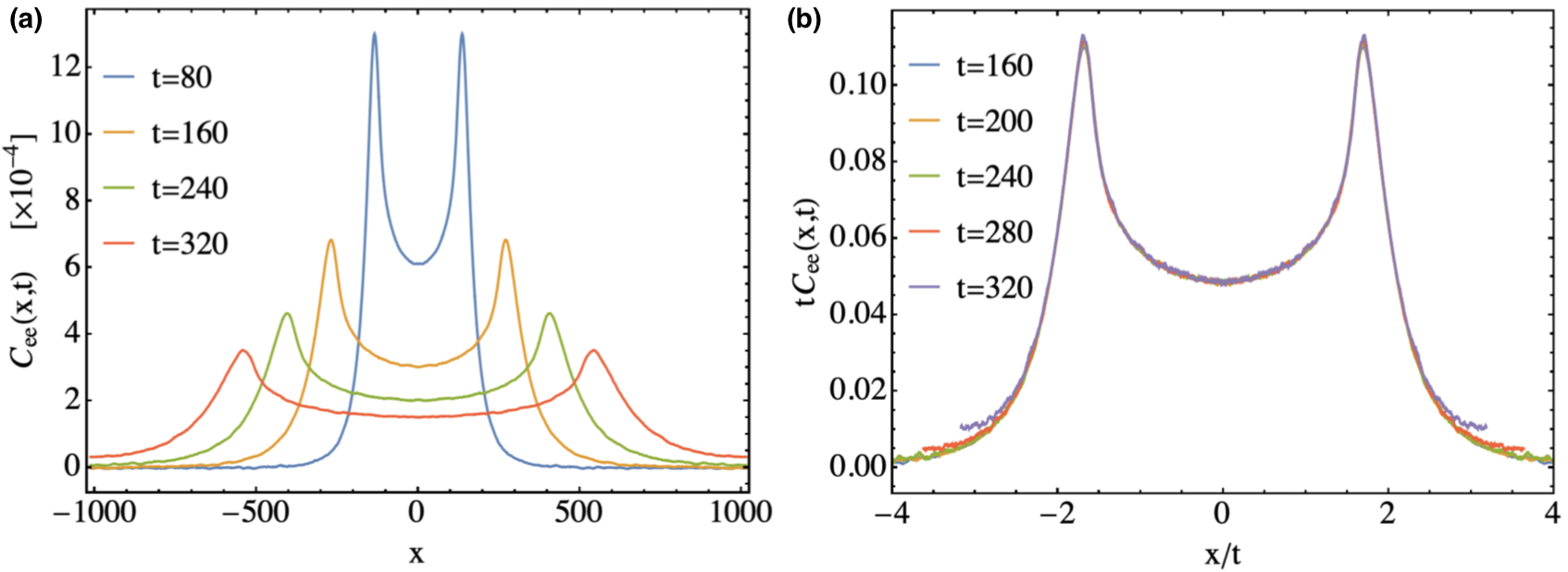}
\caption{(Isotropic regime)  Plot of the energy-energy correlation $C_{ee}(x,t)$ and the ballistic scaling of it. Parameter values: system size = 2048, final time = 320, averaging over $\sim 10^6$ initial conditions and inverse temperature $\beta$ = 1.} 
\label{fig:2}
\end{figure*}
\begin{figure*}[]
\includegraphics[width=\textwidth]{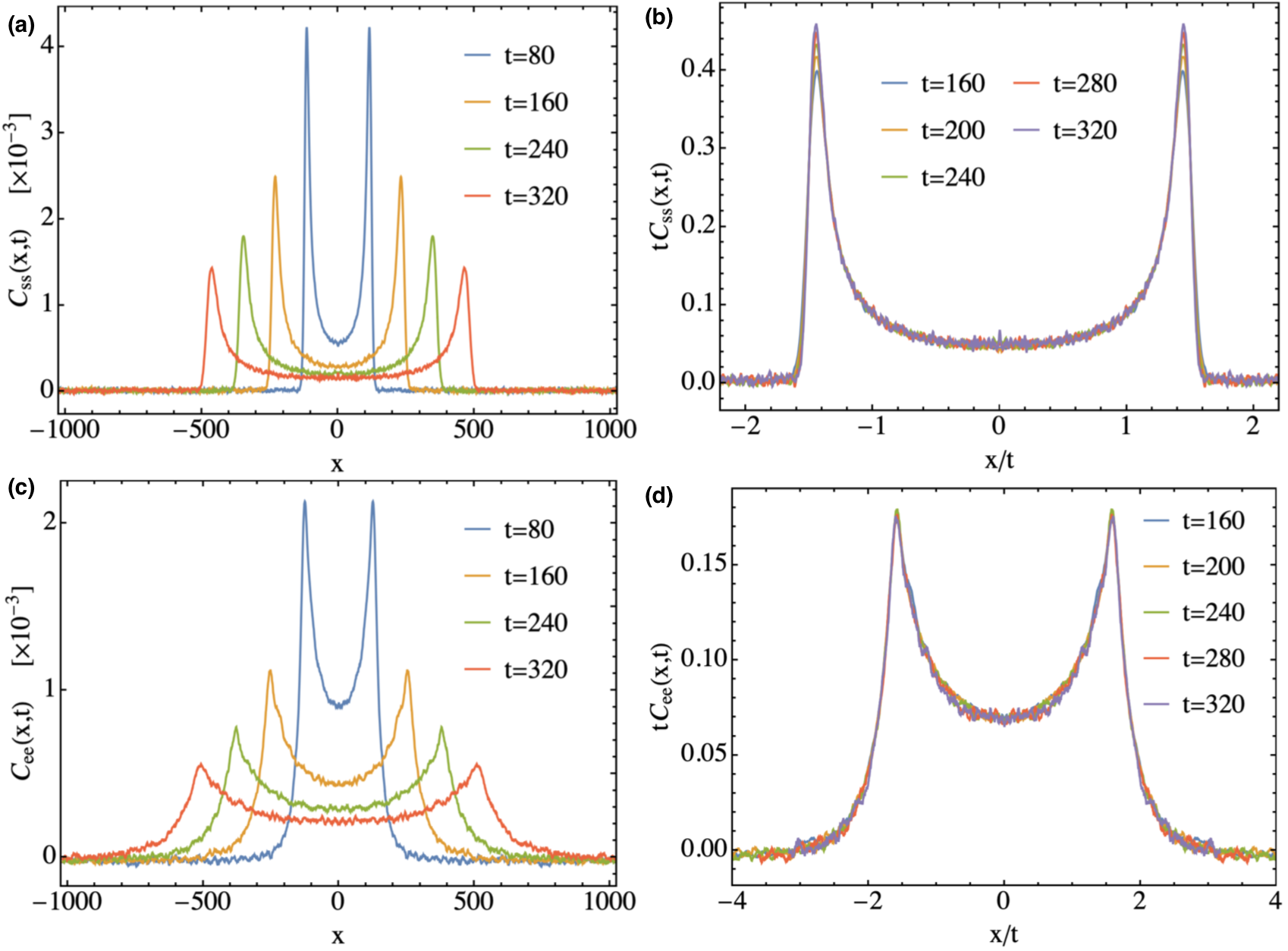}
\caption{(Easy plane regime) Plot of the spin-spin correlation $C_\mathrm{ss}(x,t)$ and energy-energy correlation $C_{ee}(x,t)$ in easy-plane regime and corresponding ballistic scalings. Parameter values: system size =$2048$,  averaging over $\sim 4\times 10^4$ initial conditions and inverse temperature $\beta=1$.}
\label{fig:3}
\end{figure*}
\begin{figure*}[]
\includegraphics[width=\textwidth]{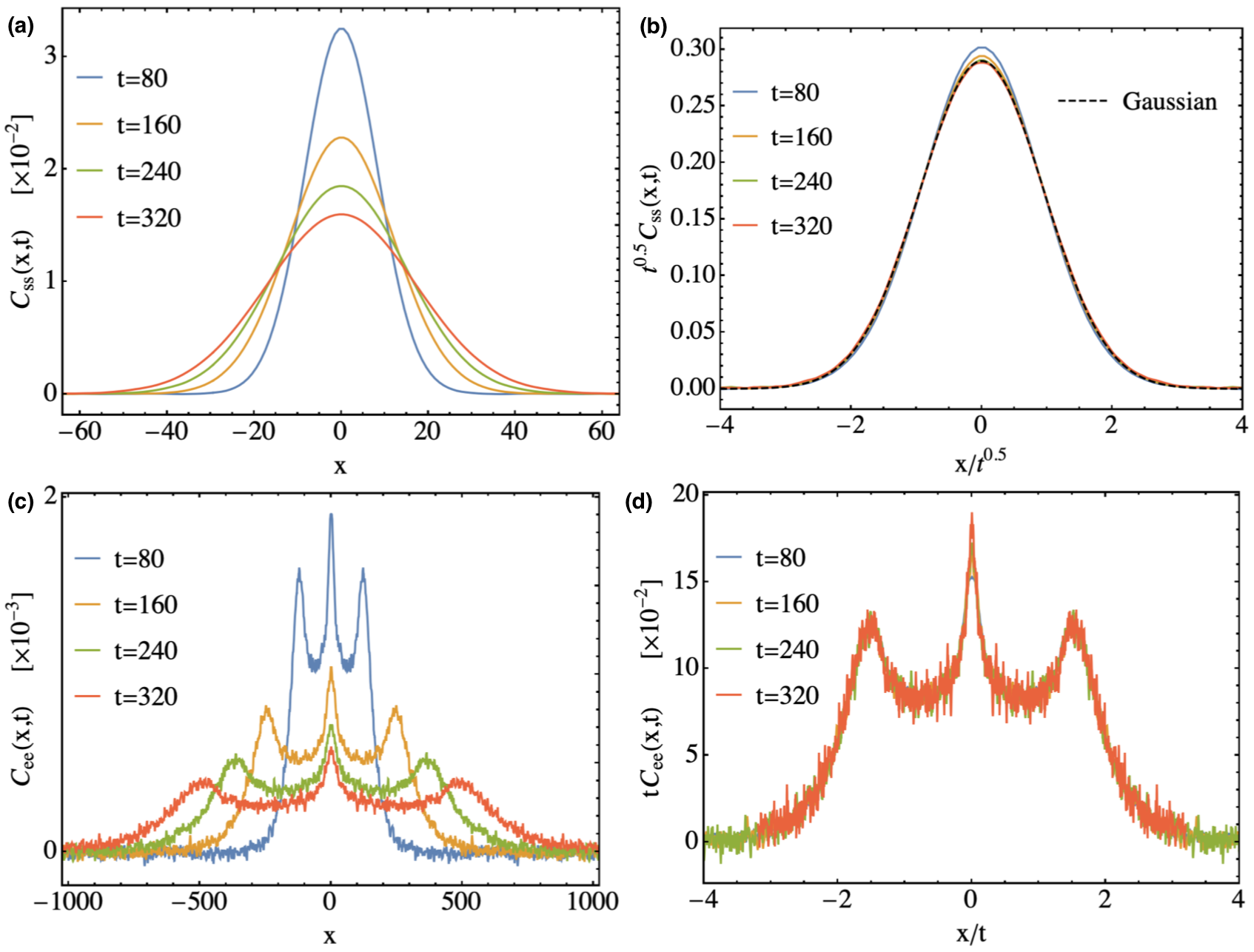}
\caption{(Easy axis regime)  Spin-spin correlation $C_\mathrm{ss}(x,t)$ and energy-energy correlation $C_{ee}(x,t)$ in easy-axis regime. In (b) we show the diffusive scaling of spin correlations while in (d) we see the  ballistic scaling of energy correlations. Parameter values: system size = $2048$, averaging over $\sim 4\times 10^4$ initial conditions and inverse  temperature $\beta=1$.}
\label{fig:4}
\end{figure*} 
\begin{figure*}[]
\includegraphics[width=\textwidth]{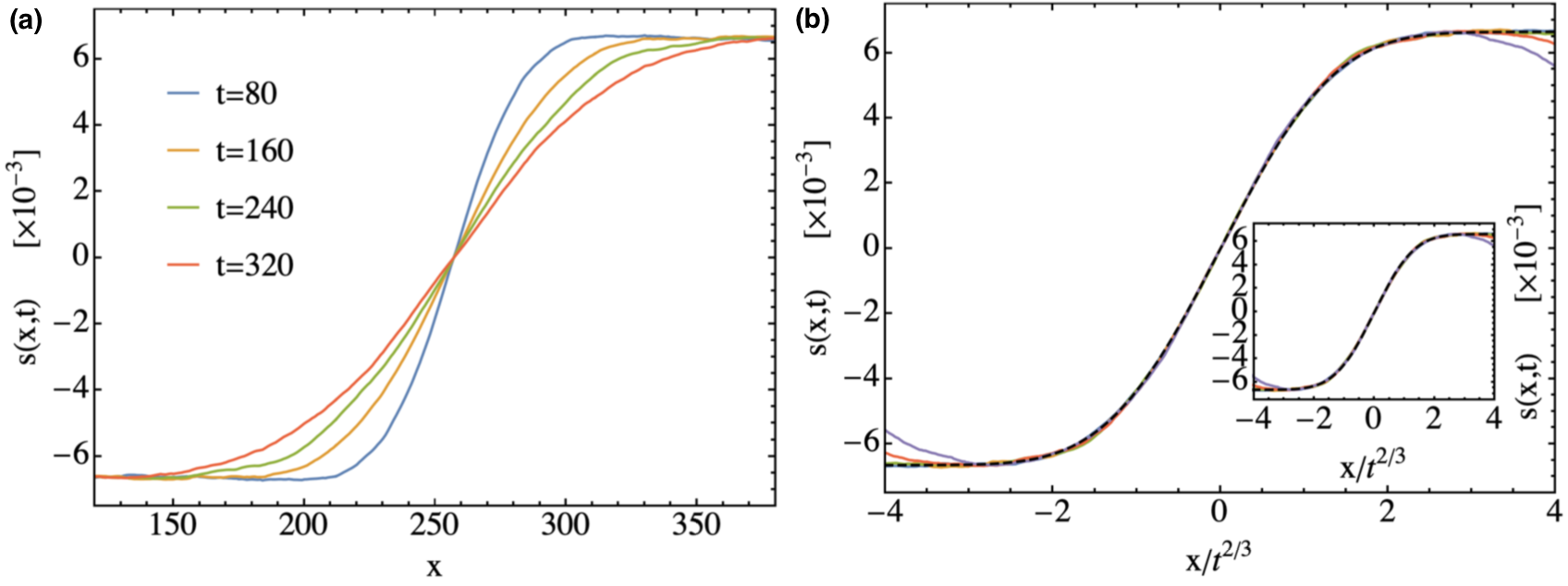}
\caption{(Isotropic regime) (a) Magnetization profile at different times starting from a step initial condition. (b) Collapse of the data under a $t^{2/3}$ scaling. The dashed line corresponds to the integrated KPZ scaling function \eqref{eq:2.11}. Inset shows the fit with integrated Gaussian, namely the Error function. Although the $2/3$ scaling is prominent, we cannot distinguish the (integrated) Gaussian and KPZ here. Parameter values: System size = $512$,  inverse temperature $\beta=1$ and averaging over $\sim 8\times 10^5$ initial conditions.} 
\label{fig:5}
\end{figure*}
We use  Metropolis Monte Carlo sampling  to generate the canonical ensemble. Starting from an ordered initial configuration, we allow $5000$ Monte Carlo swipes to make sure that the system has reached thermal equilibrium at the desired temperature. Once equilibrium has reached, we drop $500$ swipes every time we generate a new thermal configuration to use as the initial condition for the time evolution. Thereby one  ensures that the initial conditions used in the time evolution are sufficiently uncorrelated among themselves. All averages are taken over these initial conditions. The step initial profile is generated by equilibrating the system using a step magnetic field of the appropriate size at given temperature. In our study we  chose the value of $\beta=1$.  At higher temperatures, the average energy per site increases and the spins access the steeper parts of the inverted log potential [see \eqref{eq:2}] and as a result the simulation using the  adaptive step size algorithm becomes very slow [see discussion around Eq.~\eqref{eq:ES}]. For the choice $\beta=1$,  the simulation efficiency is reasonable and it is expected that our main results should be valid at other temperatures.

\section{Simulation results for equilibrium dynamical correlations} 
\label{sec:eqsim}
\subsection{Isotropic regime}
This corresponds to the choice $\rho\rightarrow 0$ in \eqref{eq:2}, which leads to the simpler form of the Hamiltonian \eqref{eq:3}. In this regime, spins have no directional bias and lie uniformly on the unit sphere. At infinite temperature these directions don't have any correlation but at finite and low temperatures the correlation grows. In Fig.~\ref{fig:1}(a) we plot the spin-spin correlation function $C_{\mathrm{ss}}(x,t)$ for $\beta=1$. We see a very good $x\sim t^{2/3}$ scaling of the data. In  Fig.~\ref{fig:1}(b) we compare the scaled data with a Gaussian distribution, while in Fig.~\ref{fig:1}(c) we compare the same data with the KPZ distribution. We first compute the sum $\sum_j C_\mathrm{ss}(j,t)$, which is independent of time and gives an estimate of the area under the fit curve. This is essentially the value of $\chi$ in \eqref{1.2}. Then we find the best fit parameter $\Gamma$ using the NonlinearModelFit function of Mathematica. In particular, we found that $\chi = 0.526698$ and $\Gamma=1.93609$ for $f_\mathrm{KPZ}$ and $1.21582$ for $f_\mathrm{Gaussian}$. Although the distinction is not so significant on this scale, we see that a much better fit is obtained with the KPZ distribution. The distinction becomes very prominent in the log plot shown in Fig.~\ref{fig:1}(d). This is because the KPZ scaling function differs from a Gaussian only in the tails. Although spin transport is superdiffusive in this regime of the Hamiltonian, energy transport is ballistic. Energy correlations are plotted in Fig.~\ref{fig:2} which show a clear ballistic scaling.

Note that, in many cases,  the diffusive or superdiffusive modes come coupled with the ballistic modes and to see them one needs to subtract the ballistic contributions, which is a difficult task in general \cite{doyon2018}. In our case it turns out that for spin transport  at the  isotropic point the ballistic contribution does not exist and we directly see the superdiffusive mode.

\subsection{Easy-plane regime}
This corresponds to the choice $\rho>0$ in Eq.~(\ref{eq:2}). Spins tend to lie near the $x-y$ plane at finite temperatures. We use the value $\rho = 1$. As shown in Fig.~(\ref{fig:3}), both spin and energy show ballistic scaling in this regime. We however observe that spin transport is slower than the energy transport. In other words,
 in Fig.~\ref{fig:3} the line shapes for spin and energy transport are distinctly different. 
 
\subsection{Easy-axis regime}
This corresponds to the choice $\rho<0$ in Eq.~(\ref{eq:2}), i.e. $\gamma$ becomes purely imaginary and the trigonometric functions become hyperbolic functions in the Hamiltonian. For our purpose, we use the value $\rho = -1$. In this regime, spins have the tendency to lie near the $z$-axis at finite temperatures. As shown in Fig.~\ref{fig:4}, we now observe that  spin correlations spread diffusively while  energy correlations spread ballistically. In this particular regime we have diffusive transport of spin. In Table.~\ref{tab-summary}, we summarize the transport properties in the ILLL chain.

\section{Magnetization profile for step initial condition}
\label{sxjt}
We consider now a chain of $N=512$ spins and prepare it at the inverse temperature $\beta =1$ using a step magnetic field as described in Sec.~\ref{model} with $h_0 = 0.01$. We average over $~8\times 10^5$ such initial conditions. The resulting step height in the magnetization is $\pm 0.00665$. These step initial conditions are  evolved according to the isotropic Hamiltonian \eqref{eq:3} and we monitor the average magnetization profile $s(x,t)$ at later times. Magnetization profiles at different times are shown in Fig.~\ref{fig:5}(a), while Fig.~\ref{fig:5}(b) shows the $2/3$ scaling of $s(x,t)$. This is expected from \eqref{eq:2.11} and our previous finding of $2/3$ scaling of $C_{SS}(x,t)$ in the isotropic regime. Although $s(x,t)$ correctly reproduces  the scaling exponent, the data is noisy and not accurate enough for us to rule out  a fit to an error function (integral of a Gaussian). In Fig.~\ref{fig:5}(b) we show the fit with integral of $f_\mathrm{KPZ}$ and, in the inset, we show the fit with the error function. Much more averaging over the initial conditions is required to arrive at smoother data shown here. Here we are essentially dealing with \eqref{eq:2.11}. This equation is supposed to be exact in the linear response limit $h_0 \rightarrow 0$, and so one should recover the same values of $\Gamma$'s and $\chi$ obtained from $C_\mathrm{ss}$ data by analyzing the step profile. However, in our simulations we have kept $h_0=0.01$ and as a result we observe slight deviations in the $\Gamma$ and $\chi$ values. Here we see $\chi = 0.665$ and $\Gamma = 1.74603$ for both KPZ and Gaussian functions.

\begin{table*}[]
\caption{
\centering
Summary of transport properties in the ILLL model for zero $z$-magnetization}
\begin{center}
\begin{tabular}{c c c}
\hline
\hline\\
\smallskip
\textbf{Regime} & \qquad\qquad\qquad \textbf{Spin transport} & \qquad\qquad\qquad \textbf{Energy transport}\\[1ex]
\hline
\hline\\
\smallskip
Easy plane $(\rho >0)$ & \qquad\qquad\qquad Ballistic & \qquad\qquad\qquad Ballistic\\[1ex]
\hline\\
\smallskip
Isotropic $(\rho \rightarrow 0)$ & \qquad\qquad\qquad
\makecell{Super-diffusive\\ scaling exponent: 2/3 \\scaling function: KPZ} &  \qquad\qquad\qquad Ballistic\\[4ex]
\hline\\
\smallskip
Easy axis $(\rho <0)$& \qquad\qquad\qquad \makecell{Diffusive \\scaling function: Gaussian} & \qquad\qquad\qquad Ballistic\\[3ex]
\hline
\hline
\end{tabular}\\[1ex]
Note: Diffusive transport implies scaling exponent = 1/2 and ballistic transport implies scaling exponent = 1.
\end{center}
\label{tab-summary}
\end{table*}

\section{Conclusions}
\label{conc}
From the study of several integrable many-body systems, there seems to be a consensus that their large scale behavior has many common features. In particular, since based on hydrodynamic type arguments, quantum models should not differ from their classical version. We presented the numerical study of the classical integrable ILLL spin chain and compared with previous studies of the quantum XXZ Heisenberg model. Our findings are summarized in Table~\ref{tab-summary} and support the view that on hydrodynamic scales classical and quantum cannot be distinguished. At the isotropic point with zero average magnetization, we find that the quantities involving spin show superdiffusive behavior with scaling exponent $2/3$ and the scaling function is KPZ. In the easy-plane regime, we find that the spin transport is ballistic, while in the easy-axis regime it is diffusive. The energy correlations are shown to exhibit ballistic scaling in all parameter regimes. To probe the KPZ behavior further we also studied the evolution of an initial magnetization step. Again, we find the $t^{2/3}$ scaling but, from these data, we  are not able to conclusively differentiate between KPZ and Gaussian scaling. 

While the numerical evidence is pointing in the expected direction, strong theoretical arguments are still missing.
Of course, a first inclination is to compare the corresponding GHD, which is available for the quantum XXZ model but currently not for its classical version. In addition, KPZ scaling requires a particular nonlinearity and noise, which is beyond conventional GHD.
 
\section{Acknowlegements}
A.D. would like to acknowledge the support from Grant No. EDNHS ANR-14-CE25-0011 of the French National Research Agency (ANR) and from Indo-French Centre for the Promotion of Advanced Research (IFCPAR) under Project No. 5604-2. M.K. gratefully acknowledges the Ramanujan Fellowship SB/S2/RJN-114/2016 from the Science and Engineering Research Board (SERB), Department of Science and Technology, Government of India. M. K. also acknowledges support from the Early Career Research Award, No. ECR/2018/002085  from the Science and Engineering Research Board (SERB), Department of Science and Technology, Government of India. M.K. would like to acknowledge support from the Project No. 6004-1 of the Indo-French Centre for the Promotion of Advanced Research (IFCPAR).  This research was supported in part by the International Centre for Theoretical Sciences (ICTS) through the program - Universality in random structures: Interfaces, Matrices, Sandpiles (code: ICTS/urs2019/01). The numerical simulations were done on \textit{Mowgli, Mario}, and \textit{Tetris} High Performance Clusters of ICTS-TIFR.

\end{document}